\documentclass[conference]{IEEEtran}
\IEEEoverridecommandlockouts
\usepackage{cite}
\usepackage{verbatim}
\usepackage{braket}
\usepackage{amsmath,amssymb,amsfonts}
\usepackage{algorithmic}
\usepackage{graphicx}
\usepackage{textcomp}
\usepackage{xcolor}
\usepackage{listings}
\def\BibTeX{{\rm B\kern-.05em{\sc i\kern-.025em b}\kern-.08em
    T\kern-.1667em\lower.7ex\hbox{E}\kern-.125emX}}
\begin{document}

\title{A physics lab inside your head: \\ Quantum thought experiments as an educational tool\\
\thanks{I thank the Heilbronn Institute for Mathematical Research and IBM Quantum for their generous support.}
}

\IEEEoverridecommandlockouts \IEEEpubid{\begin{minipage}{\textwidth}\ \\[12pt]\ \copyright2023 IEEE. Personal use of this material is permitted. Permission from IEEE must be obtained for all other uses, in any current or future media, including reprinting/republishing this material for advertising or promotional purposes, creating new collective works, for resale or redistribution to servers or lists, or reuse of any copyrighted component of this work in other works.\hfill \end{minipage}}

\author{\IEEEauthorblockN{Maria Violaris}
\IEEEauthorblockA{\textit{University of Oxford} \\
\textit{IBM Quantum}\\
Oxford, United Kingdom \\
maria@violaris.com}
}

\maketitle

\begin{abstract}
Thought experiments are where logical reasoning meets storytelling, catalysing progress in quantum science and technology. Schrödinger's famous cat brought quantum science to the public consciousness, while Deutsch's thought experiment to test the many-worlds and Copenhagen interpretations involved the first conception of a quantum computer. I will show how presenting thought experiments using quantum circuits can demystify apparent quantum paradoxes, and provide fun, conceptually important activities for learners to implement themselves on near-term quantum devices. Additionally, I will explain how thought experiments can be used as a first introduction to quantum, and outline a workshop based on the ``quantum bomb tester" for school students as young as 11. This paper draws upon my experience in developing and delivering quantum computing workshops in Oxford, and in creating a quantum paradoxes content series with IBM Quantum of videos, blogs and code tutorials.    
\end{abstract}


\section{Introduction}

Thought experiments have long played a crucial role in the progress and understanding of science. The Ancient Greek philosopher Democritus conjectured the existence of an indivisible atom, envisioning the process of continuous division of matter until reaching these fundamental units. Galileo used a thought experiment involving two spheres of different masses, demonstrating that all objects fall at the same speed regardless of their mass, and challenging Aristotle's theory of gravity. Einstein, in one of his many thought experiments, envisioned travelling alongside a beam of light, which was foundational to the development of his special theory of relativity. Now, one of the most famous thought experiments in existence is Schrödinger's cat \cite{Schrodinger1935} -- a story so interesting and accessible that it has turned the most counter-intuitive consequences of quantum physics into a well-known meme within popular culture.

Thought experiments harness the power of storytelling to provide genuine insights into quantum science, safely within an ideal lab where the equipment does not break and we can coherently control everything from bacteria to brains. While Schrödinger's cat demonstrates the problem of macroscopic superpositions, other thought experiments show other core quantum concepts: Wigner's friend shows the measurement problem \cite{Wigner1961}, Deutsch's test of many-worlds enables collapse and no-collapse theories to be distinguished \cite{deutsch1985quantum}, the double-slit shows wave-particle duality \cite{Feynman1965}, and the delayed-choice quantum eraser yields insights into decoherence and entanglement \cite{scully1982quantum, Kim2000}. Some quantum thought experiments elucidate their classical analogs: the quantum Maxwell's demon explores the work cost of quantum information erasure \cite{zurek1986frontiers, Leff2003}, while closed-timelike-curves demonstrate how quantum mechanics could solve the grandfather paradox of time-travel \cite{deutsch1991quantum}.

While thought experiments can be exceedingly powerful in catalysing new research and understanding, they can also cause confusion: many are thought of as paradoxes, due to being counter-intuitive or even in apparent contradiction with fundamental physical principles. It can be difficult to understand and explain the resolutions to apparent paradoxes that arise in quantum science, even when perfectly self-consistent solutions exist. 

\IEEEpubidadjcol

With the recent acceleration in quantum computing technology, more learners are becoming familiar with the formalism of quantum information theory and depicting quantum information processing using quantum circuits \cite{Nielsen2010}. I propose that we use quantum circuits to understand and explain quantum thought experiments: this provides new insights into the fundamental concepts behind each thought experiment; novel, interesting activities to implement on near-term quantum devices; and a compelling storytelling-based entry point to quantum information technologies for those new to the field.

In this paper, I will explain how quantum thought experiments can be expressed as quantum circuits. These form the basis of activities to create and run on near-term and current quantum devices and simulators, demonstrated using Qiskit and IBM Quantum devices \cite{Qiskit}. In particular, I will show that a range of famous quantum thought experiments, in what seem to be vastly different settings, have self-consistent explanations when measurement devices are treated as quantum systems. 

After showing a range of worked-out examples, I will summarise a general recipe for resolving new quantum thought experiments by translating them into quantum circuits. Then I will outline a quantum coding workshop for introducing quantum science and technology to beginners using the quantum bomb tester thought experiment \cite{Elitzur1993}. This workshop can fill a 1-hour long session, and be accessible for learners as young as 11, though I will offer some suggestions for adapting it to different needs.

\section{Quantum thought experiments}

First I will explain how core quantum concepts can be understood and explained using quantum thought experiments, by translating them into quantum circuits that can be run on quantum computers. The following collection of thought experiments can be used to gain a robust intuition for the implications of macroscopic superposition, connections between measurement and entanglement, tests for universality of quantum theory, wave-particle duality, decoherence, and interaction-free measurement. 

\subsection{Schrödinger's cat}

\begin{figure}[htbp!]
\centerline{\includegraphics[width=0.5\textwidth]{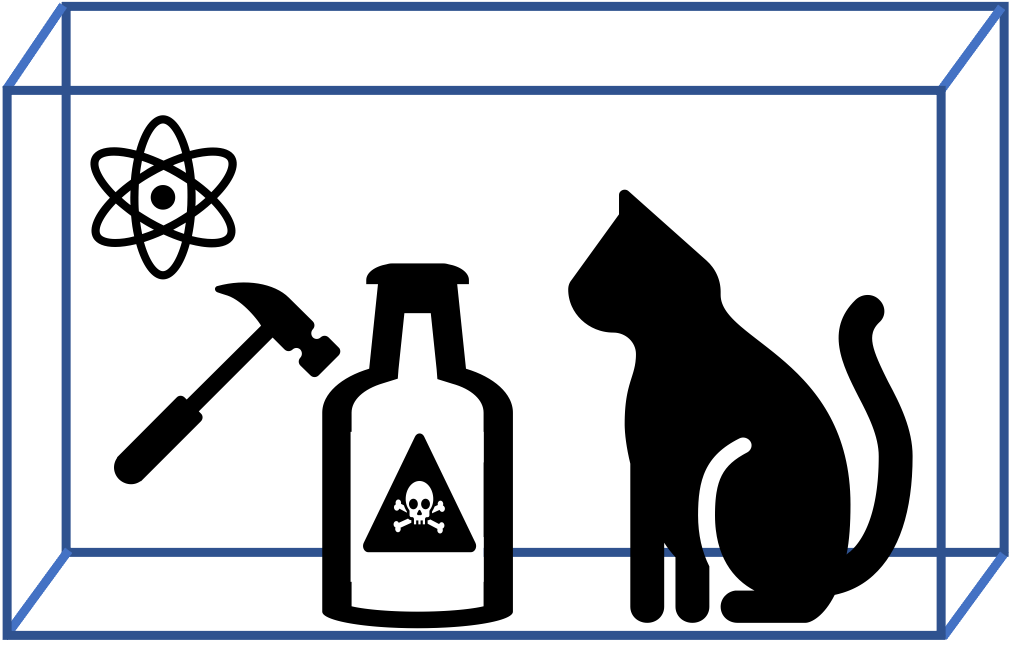}}
\caption{The Schrödinger's cat thought experiment involves a cat trapped in a box with a vial of poison, that will be smashed open by a hammer if the radioactive atom decays, killing the cat.}
\label{fig:1_cat}
\end{figure}

\begin{figure}[htbp!]
\centerline{\includegraphics[width=0.5\textwidth]{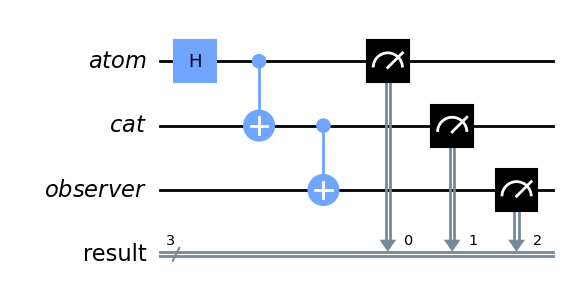}}
\caption{The quantum circuit  for a Schrödinger's cat thought experiment, when the cat and observer are modelled as quantum systems, including a Hadamard gate, CNOT gates, and Z-basis measurements. Note that the numbers on this Qiskit circuit diagram indicate which bit of the classical ``result" register the measurement outcome is stored on, not the measurement outcome itself.}
\label{fig:cat}
\end{figure}

\begin{figure}[htbp!]
\centerline{\includegraphics[width=0.5\textwidth]{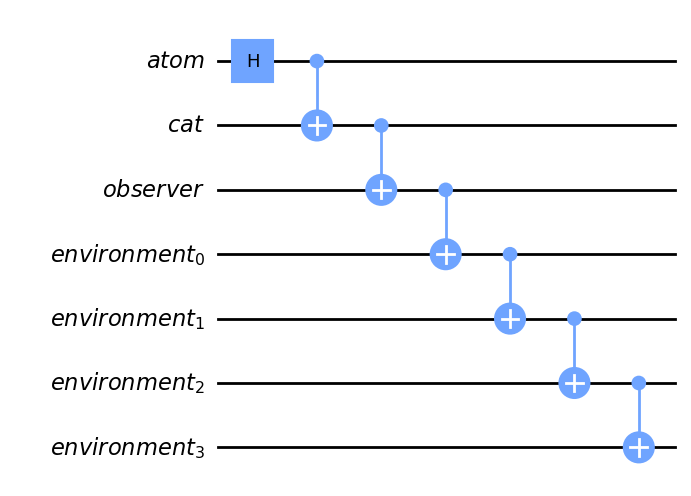}}
\caption{The quantum circuit  for a Schrödinger's cat thought experiment, where the environment is modelled as a collection of qubits, to demonstrate the process of decoherence.}
\label{fig:decohered_cat}
\end{figure}

Schrödinger's cat is a well-known thought experiment proposed by Erwin Schrödinger in 1935 \cite{Schrodinger1935}. It begins with a cat, a vial of poison, a radioactive atom, and a hammer, all inside a box (depicted in Figure \ref{fig:1_cat}). If the atom decays, the hammer is activated to break the vial, killing the cat. If the atom does not decay, the cat survives. The atom enters a superposition of decaying and not decaying. According to quantum theory, when it interacts with the hammer, poison and cat, they all become entangled. They enter a joint superposition, with one term where the atom has decayed and the cat is dead, and another where the atom has not decayed and the cat is alive.

The so-called collapse postulate of quantum mechanics says that quantum systems in superposition collapse to a single state when observed. Hence, an outside observer looking into the box forces the cat to be either dead or alive. An alternative theory is that quantum mechanics can be consistently applied to the observer, and in fact any observer or detector. Another way of phrasing this viewpoint is that quantum theory is a universal theory, so it can be applied to all parts of the universe, including observers, detectors and the environment. In this case, an observer measuring a quantum system also becomes entangled with the system, again resulting in a joint superposition of states.

To model this experiment as a quantum circuit, I will represent the radioactive atom by a qubit. It is prepared in a $\ket{+} = \frac{1}{\sqrt{2}}(\ket{0}+\ket{1})$ state by a Hadamard gate, indicating a superposition of non-decayed ($\ket{0}$) and decayed ($\ket{1}$) states. I use a second qubit to represent the cat being living ($\ket{0}$) or dead ($\ket{1}$). To model the interaction of the cat and the atom, I will introduce a CNOT gate, controlled on the atom qubit and targeted on the cat qubit. Then the cat qubit's state changes only if the atom qubit is in the decayed state ($\ket{1}$), representing the cat's death. This assumes that the cat's observation of the atom does not irreversibly collapse the system, but instead that it becomes entangled with the atom, resulting in a joint superposition state of $\ket{00}+\ket{11}$ (from this point forwards I will neglect the normalisation factor of the global state, for simplicity).

Treating the observer as a quantum system, they become entangled with the cat upon looking inside the box. I use a third qubit to denote the observer's memory, with another CNOT gate copying the cat's state to the observer's memory. Hence, the atom, cat, and observer's memory evolve into a large entangled superposition: $\ket{000}+\ket{111}$. The corresponding quantum circuit is depicted in Figure \ref{fig:cat}, created using Qiskit \cite{Qiskit}, which was used for all the circuit diagrams in this paper. The Qiskit code to create the circuit is: 

\begin{lstlisting}
from qiskit import QuantumCircuit, ClassicalRegister, QuantumRegister

atom = QuantumRegister(1, name='atom')
cat = QuantumRegister(1, name='cat')
observer = QuantumRegister(1, name='observer')
result = ClassicalRegister(3, name='result')
qc = QuantumCircuit(atom, cat, observer, result)

qc.h(atom)
qc.cx(atom, cat)
qc.cx(cat, observer)
qc.measure(atom[0], result[0])
qc.measure(cat[0], result[1])
qc.measure(observer[0], result[2])
\end{lstlisting}

Overall, the atom-cat-observer statevector evolves as follows: 
\begin{align}
\begin{aligned}
\ket{000} &\xrightarrow{} \frac{1}{\sqrt{2}}(\ket{0}+\ket{1})\ket{00}\\
&\xrightarrow{} \frac{1}{\sqrt{2}}(\ket{00}+\ket{11})\ket{0} \\
&\xrightarrow{} \frac{1}{\sqrt{2}}(\ket{000}+\ket{111})
\end{aligned}
\end{align}

The joint state contains a term where the observer saw the cat alive, and a term where the observer saw the cat dead. Then, why does the observer perceive only a single outcome? This can be attributed to decoherence \cite{zurek2003decoherence} -- as the observer interacts with the rest of the environment, more and more of the environment joins in the entangled superposition. The environment can be modelled as an array of qubits, with a series of CNOT gates transmitting information from the observer to each environment qubit. After decoherence, the atom, cat, observer and environment are in a joint state $\ket{00...0}+\ket{11...1}$. The quantum circuit that includes decoherence from the environment is depicted in Figure $\ref{fig:decohered_cat}$.

Note that whether or not macroscopic superpositions can be realized remains an active topic of research amongst the quantum community, with recent experiments putting increasingly macroscopic systems into superposition (e.g. \cite{bild2023cat}). What the quantum circuit representation of Schrödinger's cat helps to demonstrate is that there is no internal contradiction in the existence of macroscopic superpositions. 

\subsection{Wigner's friend}

\begin{figure}[htbp!]
\centerline{\includegraphics[width=0.5\textwidth]{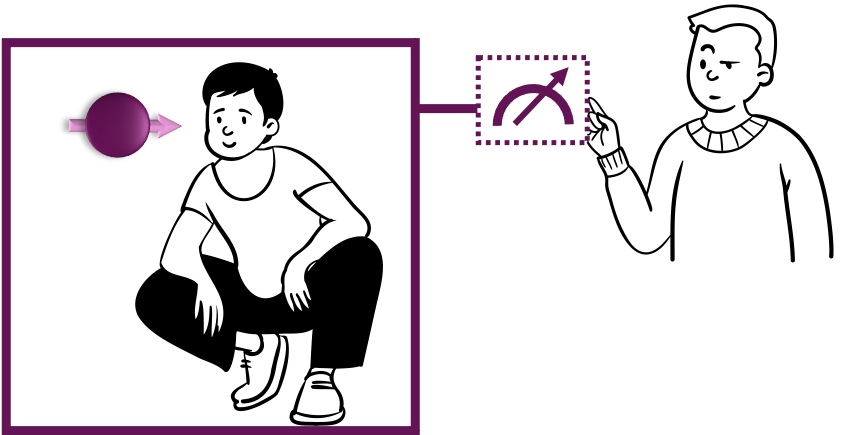}}
\caption{In the Wigner's friend thought experiment, Wigner's friend (left) measures a quantum system that begins in a superposition of two states. Then, Wigner (right) asks his friend what result he got, which is a measurement of his friend's result.}
\label{fig:2_friend}
\end{figure}

\begin{figure}[htbp!]
\centerline{\includegraphics[width=0.5\textwidth]{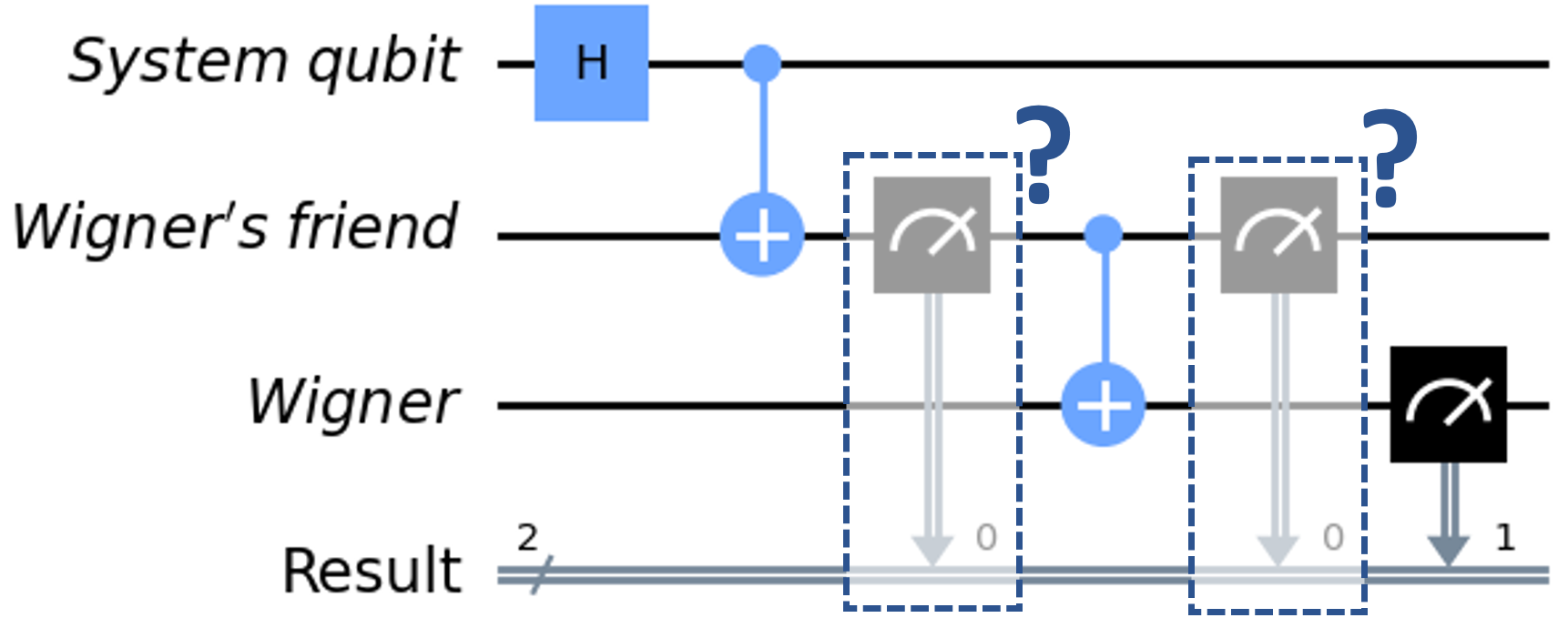}}
\caption{A quantum circuit for the Wigner's friend thought experiment, with the two cases of Wigner's friend doing an irreversible measurement before Wigner's measurement, and Wigner doing the irreversible measurement, represented by including either the mid-circuit measurement before the final CNOT (left box with dashed border) or measurement after the CNOT (right box with dashed border) respectively.}
\label{fig:wigner}
\end{figure}

The Wigner's friend thought experiment explores the nature of measurement in quantum mechanics \cite{Wigner1961}. In this scenario, Wigner's friend observes a quantum system in a superposition, which I will represent as a qubit.
The friend perceives a definite state, either $\ket{0}$ or $\ket{1}$, which suggests a projection of the qubit to a single state upon measurement.

However, Wigner begins isolated from his friend and the qubit, such that from Wigner's perspective, his friend and the qubit just became entangled. This entanglement resulted in a joint superposition of his friend observing $\ket{0}$ and the qubit being $\ket{0}$, and the converse state. The setup is depicted in Figure \ref{fig:2_friend}.

To visualize this in a quantum circuit, I will model the friend's memory as a qubit. I first apply a Hadamard gate to the qubit that will be measured, preparing it in a superposition of $\ket{0}$ and $\ket{1}$. Then I represent the friend's measurement of the qubit as a CNOT gate, which copies the qubit's state to the friend's memory, leading to a maximally entangled Bell state.

When Wigner asks his friend for the measurement result, his friend communicates either 0 or 1. According to Wigner, this instant triggers the projection of the qubit and friend into a single state. This step is illustrated by another CNOT gate, this time between the friend's memory and Wigner's memory. The resulting quantum circuit is demonstrated in Figure $\ref{fig:wigner}$, with only the measurement in the dashed box on the right included.

The crucial question then is, when did the actual projection occur? Here there is a paradox, if observation induces an irreversible collapse. In that case, the friend insists the collapse happened when the qubit was measured, whereas Wigner insists it happened when he learnt the qubit's state from his friend. Specifically, Wigner's friend would say the true quantum circuit includes the measurement before the final CNOT in Figure $\ref{fig:wigner}$. This discrepancy is known as the measurement problem.

The prevalent ``Copenhagen interpretation" suggests that a quantum system collapses into a single state upon observation. Since this explanation lacks a clear mechanism describing the measurement process, it fails to resolve the Wigner's friend paradox. This makes the Copenhagen interpretation self-contradictory, unless it is augmented with a modified explanation for measurement. 

A plausible resolution to the paradox is treating all observers, including Wigner and his friend, as quantum systems, taking seriously the depiction of their memory states as qubits in our quantum circuit. This perspective reveals that both observers encountering a distinct measurement outcome is self-consistent, since they each join an entangled superposition upon interacting with the system or each other. Therefore, considering Wigner and his friend as quantum systems resolves the paradox.

\subsection{Many-worlds vs Copenhagen interpretation test}

\begin{figure}[htbp!]
\centerline{\includegraphics[width=0.5\textwidth]{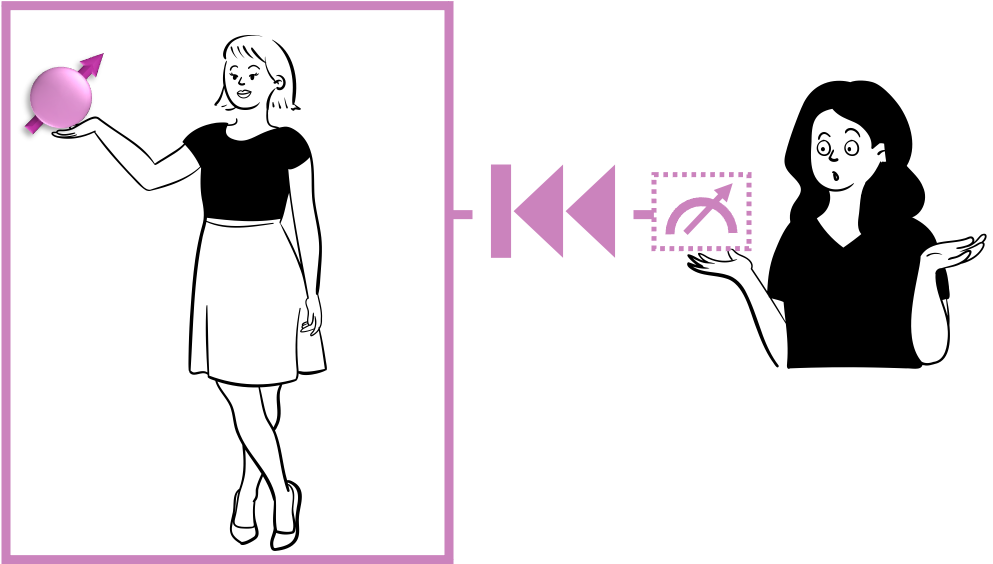}}
\caption{The many-worlds vs Copenhagen interpretation thought experiment considers a simulation running on a quantum computer, of an observer measuring a qubit in superposition (left). An outside observer (right) then reverses the measurement, after receiving confirmation that a measurement was indeed made, and checks if the qubit returned to a superposition.}
\label{fig:3_AI}
\end{figure}

\begin{figure}[htbp!]
\centerline{\includegraphics[width=0.5\textwidth]{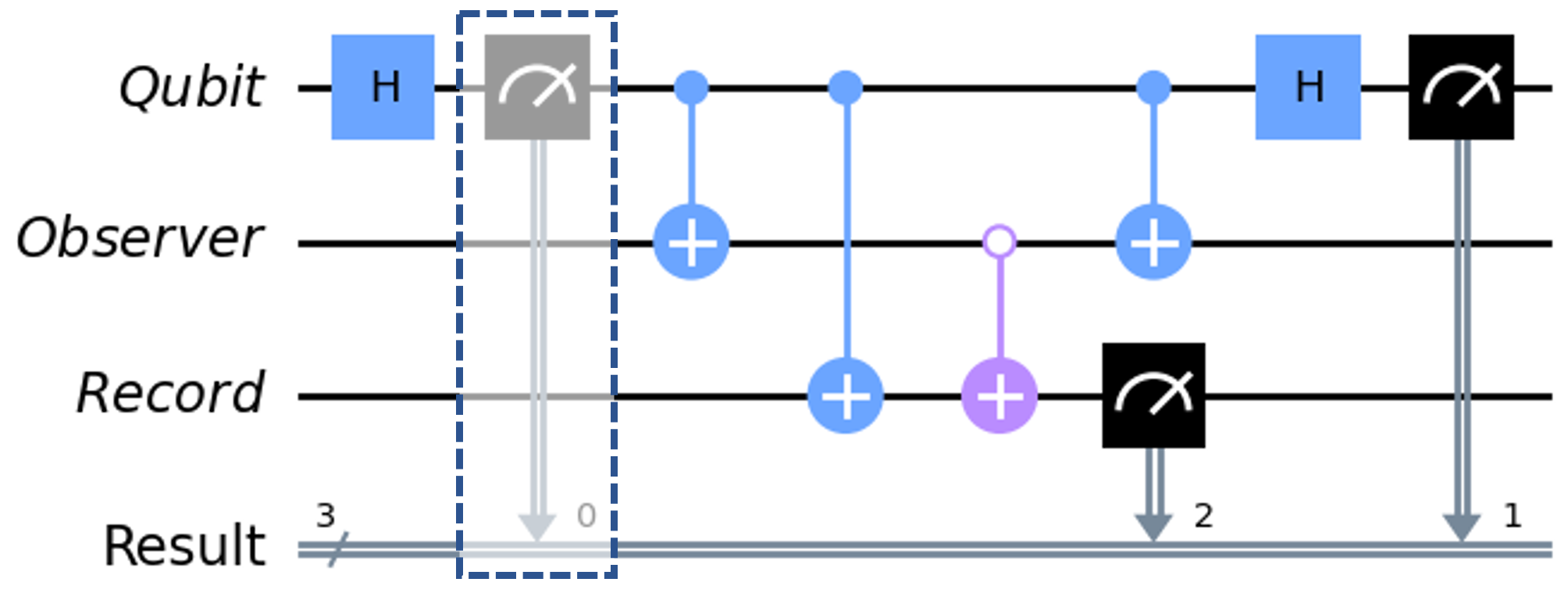}}
\caption{A quantum circuit representing the many-worlds vs Copenhagen thought experiment. The case where the simulated observer's measurement of the qubit causes an irreversible collapse is implemented by including the mid-circuit measurement in the dashed box. This measurement is omitted when the simulated observer's measurement is reversible, just implemented by an entangling CNOT gate. The circuit includes an anti-controlled NOT gate, which has an unfilled circle on the control qubit.}
\label{fig:quantum_ai}
\end{figure}

In 1985, David Deutsch devised a thought experiment to test whether quantum measurements are reversible, thereby avoiding collapse, or irreversible, inducing collapse \cite{deutsch1985quantum}. Motivated as a test for the many-worlds theory of quantum mechanics against the Copenhagen interpretation, the thought experiment also included the first description of a quantum-coherent universal quantum computer.

The thought experiment involves simulating an observer measuring a quantum system in superposition, then attempting to reverse the measurement to ascertain if the quantum system reverts to its initial superposition (depicted in Figure \ref{fig:3_AI}). Modelling the system as a qubit: if the observation caused an irreversible collapse to either $\ket{0}$ or $\ket{1}$, the measurement cannot be reversed. Conversely, if the observation is a quantum interaction resulting in entanglement, the measurement can be reversed, and the qubit returns to the original superposition of $\ket{0}$ and $\ket{1}$.

My quantum circuit representation of this thought experiment uses three qubits: one is the quantum system to be measured, beginning in the $\ket{+}$ state; the second represents the observer's memory; and a third provides a permanent record of the fact the observer made the measurement. I apply a Hadamard gate to the first qubit to prepare it in the $\ket{+}$ state, before it is measured by the observer.

For the no-collapse scenario, the observer, considered a quantum system, becomes entangled with the qubit upon measurement. This interaction is modeled by a CNOT gate, which copies the qubit's state onto the observer's memory. 

Next is arguably the ingenious part of this thought experiment. A ``record" qubit is introduced to maintain evidence that the observer indeed made a measurement, but crucially, this can be done without revealing the observer's measurement outcome. This is what will enable us to later reverse the observer's measurement, while maintaining our record of the measurement taking place. (If the record qubit stored any information about the observer's measurement outcome, then the measurement would be impossible to reverse by manipulating the observer and their qubit alone - they will have decohered due to their environment having a record of the result.)

To store this record of the fact a measurement took place, without revealing the actual information, we can implement a parity check between the system qubit and observer, targeted on the record qubit. This checks whether the measurement was done successfully by indicating whether or not the system qubit and observer are in the same state. It can be implemented using a CNOT gate and anti-CNOT gate: if the states of the observer and qubit are correlated, then the measurement was done successfully, and the Record qubit flips to $\ket{1}$; if not, it remains at $\ket{0}$.

Now to reverse the observer's measurement, we need to reverse the entanglement of the observer and the qubit. This process would require extraordinary coherent control over a living system. A more feasible but still demanding approach would be to simulate an observer on a computer, under quantum coherent control. This was the insight that led to Deutsch's description of this many-worlds vs Copenhagen thought experiment becoming the first proposal for a quantum-coherent universal quantum computer.

To reverse the entanglement of the observer and qubit in the quantum circuit, I apply a CNOT gate between the system qubit and observer. Then a final Hadamard gate on the qubit transforms its state deterministically to 0.  

If successful, the circuit simulation should yield a deterministic outcome of 0 for the qubit and 1 for the Record, indicating that a measurement did indeed take place, yet the measurement was reversed, requiring a no-collapse explanation. The overall quantum circuit is depicted in Figure $\ref{fig:quantum_ai}$, when the mid-circuit measurement in the dashed box is not included.

To represent collapse, an irreversible mid-circuit measurement is introduced before the measurement CNOT gate, causing the qubit to collapse into a single state and lose its coherence. In this case, the final gates cannot reverse the measurement and return the qubit to the $\ket{0}$ state. The final CNOT still erases the observer's memory of their observed outcome, but the system qubit remains in a fixed state of $\ket{0}$ or $\ket{1}$, not a superposition. Hence, the final Hadamard gate places the qubit in a superposition state of $\ket{+}$ or $\ket{-}$, resulting in a 50-50 chance of measuring $\ket{0}$ or $\ket{1}$ over multiple runs. This scenario's quantum circuit is displayed in Figure $\ref{fig:quantum_ai}$ with the inclusion of the mid-circuit measurement in the dashed box.

\subsection{Double slit}

\begin{figure}[htbp!]
\centerline{\includegraphics[width=0.5\textwidth]{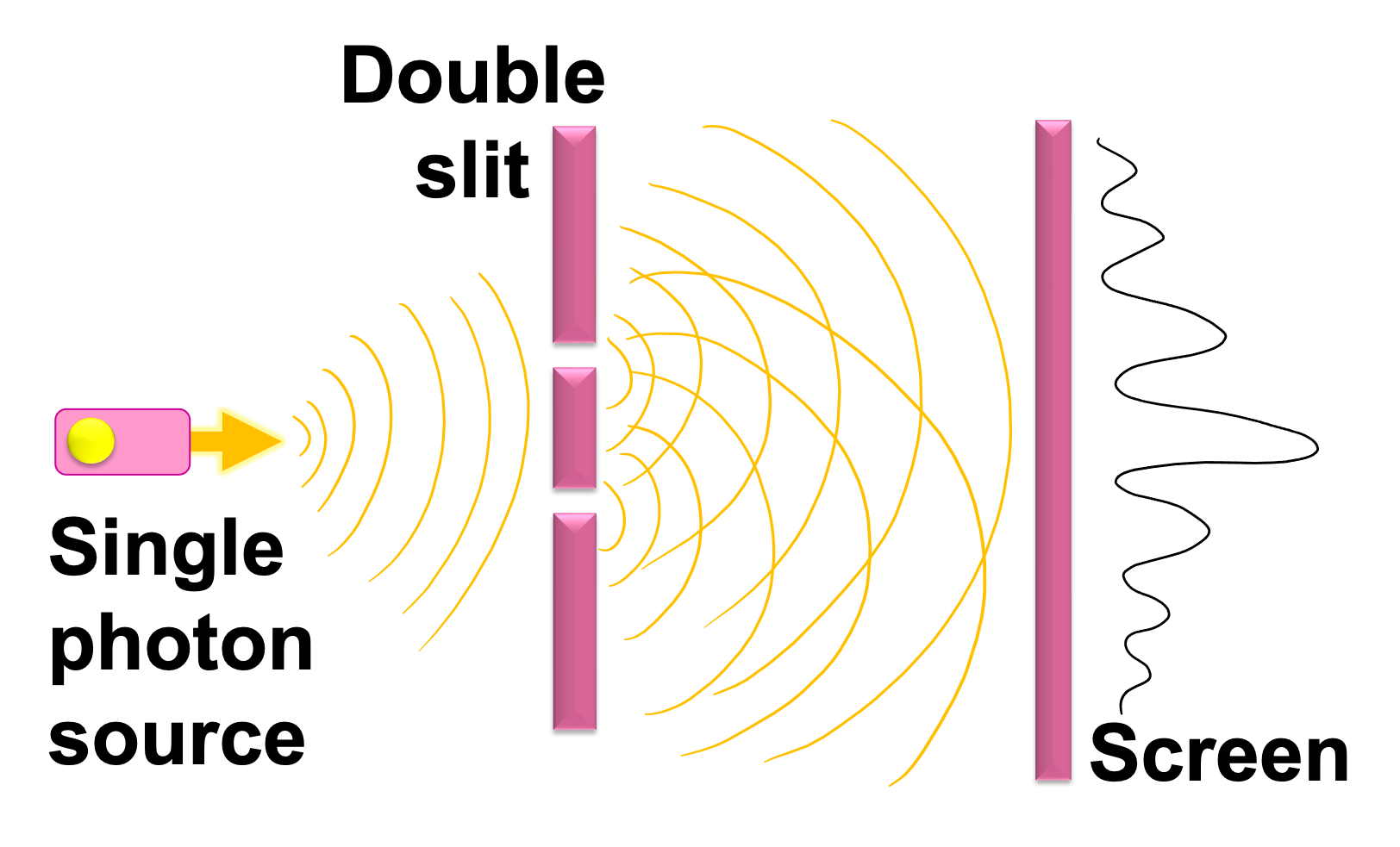}}
\caption{In the double-slit thought experiment, a single photon passing through two slits is able to interfere with itself, such that the pattern on the screen forms bright and dark interference fringes.}
\label{fig:4_double}
\end{figure}

\begin{figure}[htbp!]
\centerline{\includegraphics[width=0.5\textwidth]{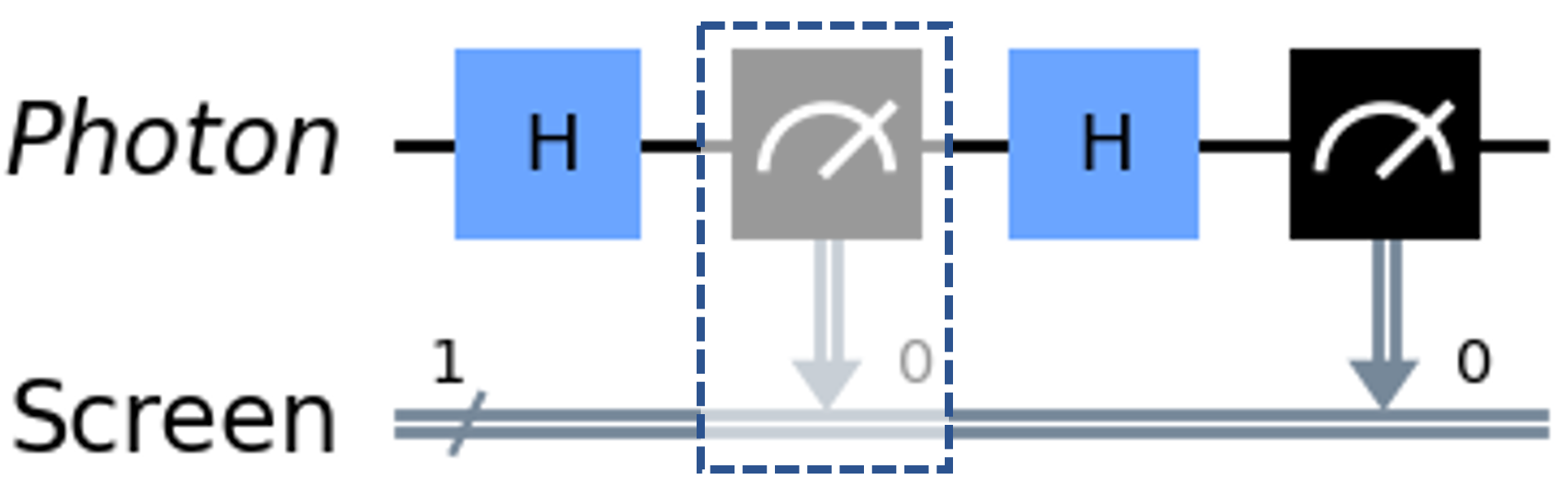}}
\caption{The quantum circuit analog for demonstrating wave-like behaviour of quantum systems does not include the measurement in the dashed box: there are only two Hadamard gates, showing how the photon can be mapped to a superposition of $\ket{0}$ and $\ket{1}$, then interfere with itself to be mapped back to a $\ket{0}$ state. The circuit for particle-like behaviour includes the mid-circuit measurement between the two Hadamard gates, extracting information about the state of the photon, which prevents it from interfering with itself at the second Hadamard gate. Hence, it does not return deterministically to the $\ket{0}$ state.}
\label{fig:double_slit}
\end{figure}

\begin{figure}[htbp!]
\centerline{\includegraphics[width=0.5\textwidth]{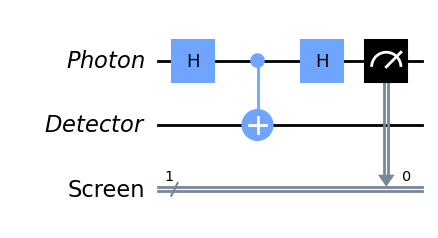}}
\caption{A quantum circuit demonstrating how the particle-like behaviour of quantum systems is not fundamentally due to classical measurements, but due to any interaction with another system (such as a detector qubit) that copies the information about its state.}
\label{fig:double_slit_decohered}
\end{figure}

The double-slit thought experiment is the standard example of so-called wave-particle duality in quantum physics \cite{Feynman1965}. Here, single quantum systems, such as individual photons or electrons, demonstrate wave-like interference. However, the interference effect disappears if we observe the particle's path. This property is often used as the first illustration of the peculiarities of quantum mechanics; in the UK, wave-particle duality forms part of some secondary school physics syllabuses. 

The conventional explanation presents this thought experiment as somewhat inexplicable: somehow quantum systems can behave both as particles and waves, making them exotic and perhaps even indescribable. By extracting the key features of the double slit thought experiment and translating it to quantum circuits, I will provide an intuitive and non-contradictory explanation for the wave- and particle-like behaviour of quantum systems, including the subtle effect of observation. 

The typical setup of a double-slit thought experiment involves directing individual photons towards a barrier with two slits. Beyond the barrier, there is a screen detecting the photons. If the photons behaved like classical particles, we would expect two bright bands on the screen, one corresponding to each slit. However, the actual result is an interference pattern of light and dark bands, like for a wave passing through the two slits, depicted in Figure \ref{fig:4_double}.

This interference pattern implies that each photon is somehow passing through both slits simultaneously and interfering with itself. If we add a detector at the slits to determine which slit the photon took, the interference pattern disappears, leaving only two bright regions corresponding to each slit, demonstrating the particle-like behavior of photons.

To visualize this better, we can use the Mach-Zehnder interferometer, an analogous experiment with a simple quantum circuit representation. This setup splits a single photon into a superposition of two paths, mirrored in a quantum circuit by a qubit, where the state $\ket{0}$ corresponds to the bottom path, and the state $\ket{1}$ corresponds to the top path. The beam-splitter creating this superposition is represented by a Hadamard gate.

\textbf{Case 1: Without a detector}

The photon encounters another beam-splitter after following both paths, where it constructively interferes with itself, merging the two parts of the superposition back into a single transmitted photon. In the quantum circuit representation, the second beam-splitter is another Hadamard gate, taking the qubit from the $\ket{+}$ state to the state $\ket{0}$. This interference can be observed by a measurement on the photon qubit, which always returns $\ket{0}$ (corresponding to a transmitted photon detected after the second beam-splitter). The quantum circuit for this wave-like behaviour is shown in Figure $\ref{fig:double_slit}$, when the measurement in the dashed box is omitted.

\textbf{Case 2: With a path detector}

If we add a detector in one of the photon's paths, it projects the photon into a definite path, preventing it from interfering with itself. At the second beam-splitter, the photon then splits into an equal superposition, with each detector having a 50$\%$ chance of activation. This is modeled in the quantum circuit by adding a measurement operation after the first Hadamard gate, yielding a 50/50 chance of ending in the state $\ket{0}$ or $\ket{1}$. The quantum circuit for this particle-like behaviour is shown in Figure $\ref{fig:double_slit}$, when the measurement in the dashed box is included.

\textbf{Case 3: Using a detector qubit}

One way to demystify the detector's impact on the photon's behavior is by modeling the detector as a quantum system. In the quantum circuit, we represent the detector as another qubit, and the detection process as a CNOT gate between the photon and detector. This entangles the photon qubit with the detector qubit. Consequently, the photon cannot interfere with itself anymore. The quantum circuit for the experiment with a detector qubit is depicted in Figure $\ref{fig:double_slit_decohered}$, reconciling the wave-like and particle-like quantum circuits by demonstrating that the key difference is the introduction of the CNOT gate from the photon qubit to another system. 

Even without ``classically" observing the detector qubit's state, just entangling it with the photon is sufficient to disrupt the interference. This demonstrates that any interaction which copies the photon's path information into the environment is enough to alter its behavior. This is the mechanism behind decoherence - it arises due to unwanted interactions between quantum systems. There is no need for any notion of a classical or irreversible measurement to explain decoherence. 

To summarize, the double-slit experiment is the standard example of how quantum systems display both wave-like and particle-like behaviors, depending on whether their path information is accessible. Understanding how the particle-like behaviour can arise from entanglement with a single detector qubit unifies these two behaviours of quantum systems, and demonstrates the mechanism behind decoherence.

\subsection{Delayed-choice quantum eraser}

\begin{figure}[htbp!]
\centerline{\includegraphics[width=0.5\textwidth]{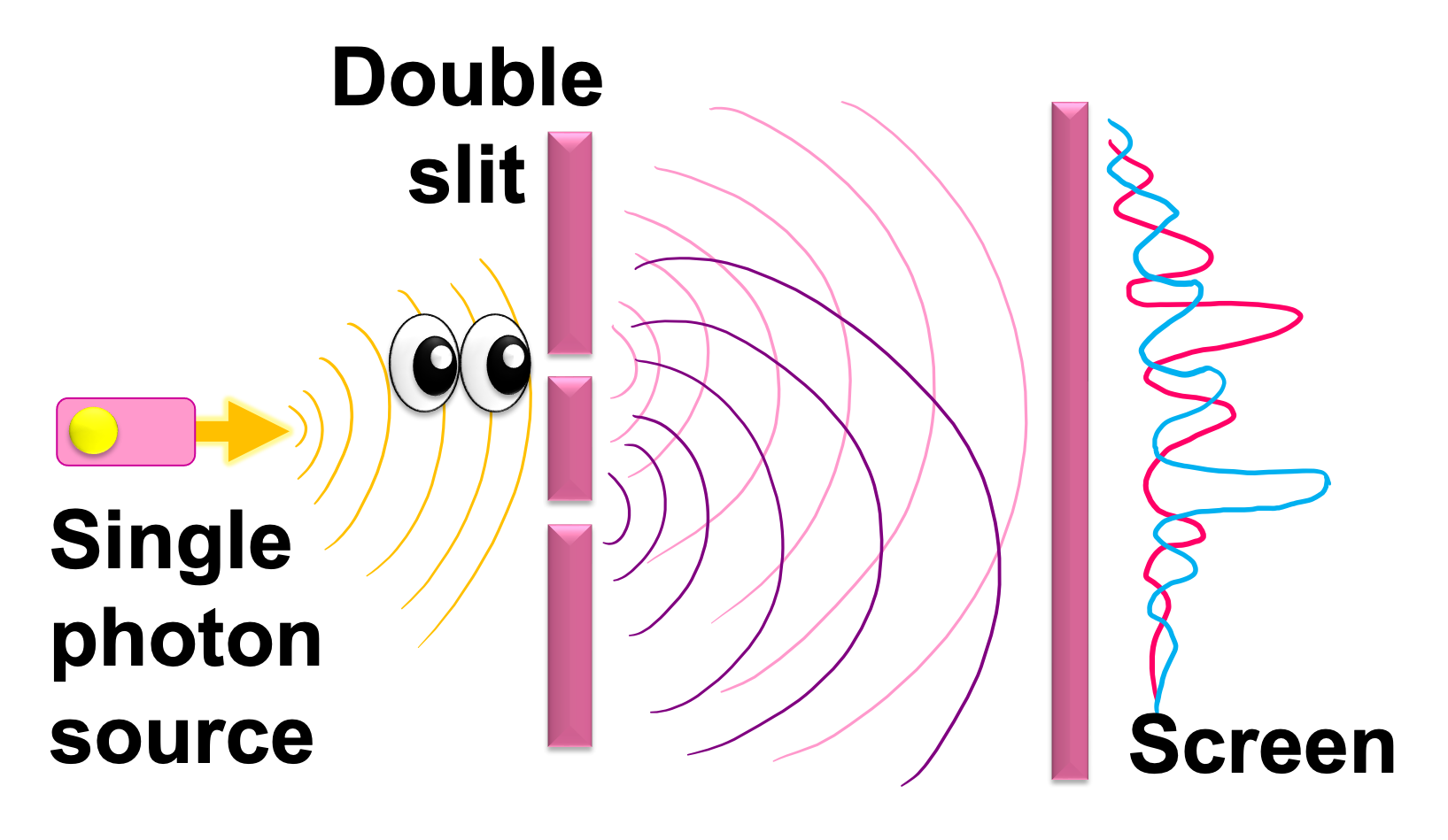}}
\caption{In the quantum eraser thought experiment, a detector qubit is used to store information about which slit the photon passed through, causing the particle-like pattern of two peaks on the screen. By later measuring the detector qubit in a different basis, and separating the patterns on the screen according to the two possible outcomes, two individual interference patterns can be extracted.}
\label{fig:5_eraser}
\end{figure}

\begin{figure}[htbp!]
\centerline{\includegraphics[width=0.5\textwidth]{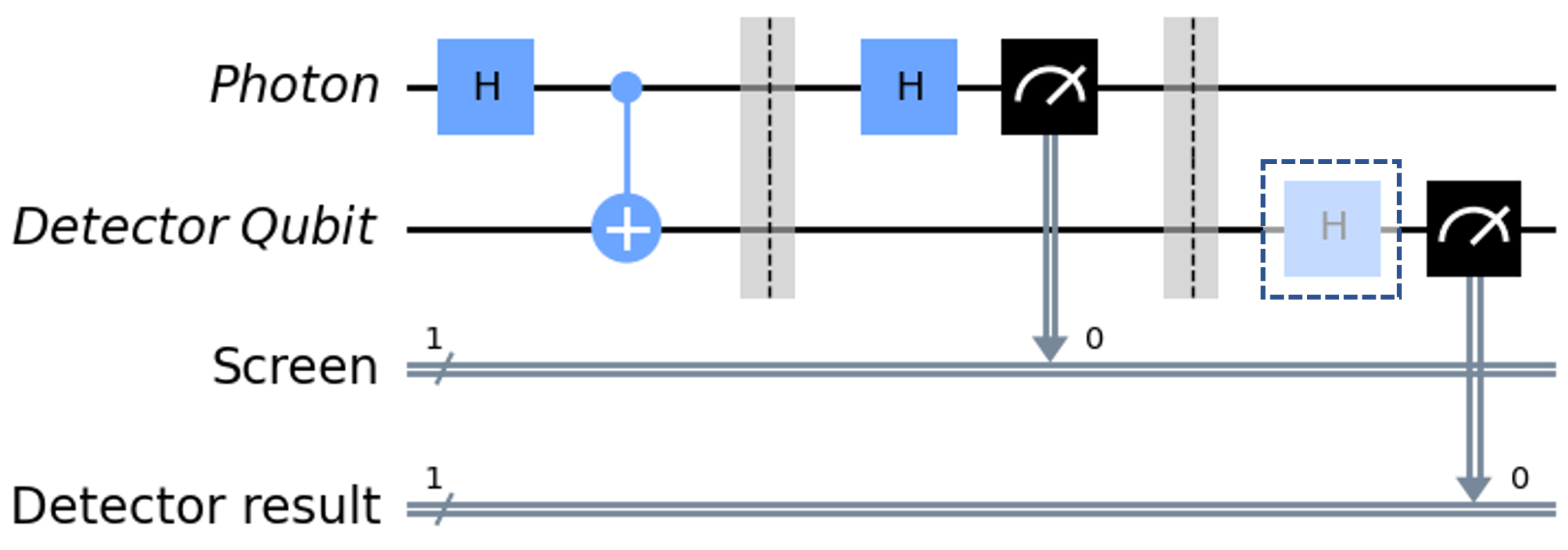}}
\caption{A quantum circuit for the quantum eraser thought experiment. By omitting the Hadamard gate in the dashed box, the detector qubit is measured in the Z-basis, extracting the information about which state the photon was in, in between the Hadamard gates. When the Hadamard in the dashed box is included, the detector qubit is measured in the X-basis. This erases information about which state the photon was in, in between the Hadamard gates, however the information extracted is correlated with the state of the photon on the screen due to entanglement.}
\label{fig:eraser}
\end{figure}

The delayed-choice quantum-eraser experiment is a sophisticated adaptation of Wheeler's delayed-choice experiment, which itself extends the classic double-slit experiment \cite{scully1982quantum, Kim2000}. It seems to hint that present-time measurements of a quantum system could retroactively determine its past behavior -- specifically, whether it acted as a wave or a particle. 

Consider the setup of the double-slit thought experiment. As in the previous section, instead of employing a classical detector to discern the slit a photon passed through, we use a small, controlled quantum system, or a ``detector qubit", that interacts with the photon at the slits. 

The key insight for this thought experiment is that the detector qubit can be measured in a way that reveals which slit the photon went through, or it can be measured such that this information is lost (i.e. erased). This measurement choice can be made at any time after the detection of the photon on the screen, even years later.

The key question is then, what kind of pattern should we anticipate on the screen of the double-slit experiment? I explained using the double-slit experiment quantum circuit, that mere interaction of the detector qubit with the photon prevents the typical interference pattern from emerging. This makes intuitive sense, since when which-path information can be extracted, interference should not be possible, even if we do not yet copy this information into our macroscopic measurement instruments. 

The more subtle part arises when the detector qubit is measured in a way that erases the information about the photon's path. Using the results of this measurement, it is possible to discern an interference pattern from the screen data. Suppose our detector qubit yields two outcomes, $\ket{0}$ or $\ket{1}$. If we separate the photon impacts on the screen corresponding to a $\ket{0}$ from those corresponding to a $\ket{1}$, interference patterns emerge in the two distinct sets of patterns on the screen.

Therefore, if we choose to measure our detector in a way that erases the photon's path information, we can extract interference patterns. Naively, this seems to suggest that individual photons interfered with themselves and passed through both slits, leading to a perplexing paradox. It appears as if we can retroactively dictate the photon's behavior (whether it behaved as a particle through one slit, or as a wave through both slits) by choosing how to measure our detector qubit after the photons hit the screen. This implies a bizarre ability to influence a photon's past based on current actions. 

To resolve this paradox, we can again represent it as a quantum circuit, as a variation of the final quantum circuit in the double-slit discussion in Figure $\ref{fig:double_slit_decohered}$. We have two options: the first is to apply a standard Z-measurement to the detector qubit, resulting in a random mix of outcomes revealing which path the photon took, shown in Figure $\ref{fig:eraser}$ when omitting the Hadamard in the dashed box. The second is to make an X-measurement (by applying a Hadamard gate and then a Z-measurement), which retrieves no path information, shown in Figure $\ref{fig:eraser}$ with the Hadamard in the dashed box included.

Let's see how to extract the apparent interference pattern from the X-measurement results to resolve the paradox. We find that the X-measurement outcomes on the detector qubit are exactly correlated with the measurement outcomes of the photon. We can see from the quantum circuit that this is exactly as expected: we prepared a Bell state between the photon qubit and detector qubit, making them maximally entangled, then measured them both in the X basis, hence their outcomes will be precisely correlated. 

In terms of the Mach-Zehnder interferometer, this means that one set of qubit detector outcomes is always correlated with one of the final photon detector outcomes - i.e. each individual run of one set of the qubit detector outcomes gives the same behaviour at the interferometer as single-photon intereference does, namely a consistent detection at just one of the detectors after the second beam-splitter. Translating this back to the double-slit context: when we isolate one set of detector outcomes and look at the screen outcomes correlated with those, we see the same behaviour as the single-photon interference setup, with bright and dark fringes. With our quantum circuit representation, we can see that this behaviour is in fact a signature of quantum entanglement between the photon and detector qubit, and it does not come from the photon interfering with itself. This insight resolves the apparent paradox: the interference pattern we are able to extract with the pathway-erasing measurements on the detector is not a signature of single-photon interference, but a signature of entanglement. 

In summary: the photon's behavior does not change based on how we measure the detector qubit. The detector qubit causes decoherence and destroys the single-photon interference pattern. Measuring the detector qubit in the X basis does not alter the photon's past behavior, but demonstrates the entanglement between the detector and photon qubits. Extracting the photon outcomes correlated with one set of detector outcomes gives the illusion of retroactively extracting a single-photon interference pattern.

\subsection{Quantum bomb tester}

\begin{figure}[htbp!]
\centerline{\includegraphics[width=0.5\textwidth]{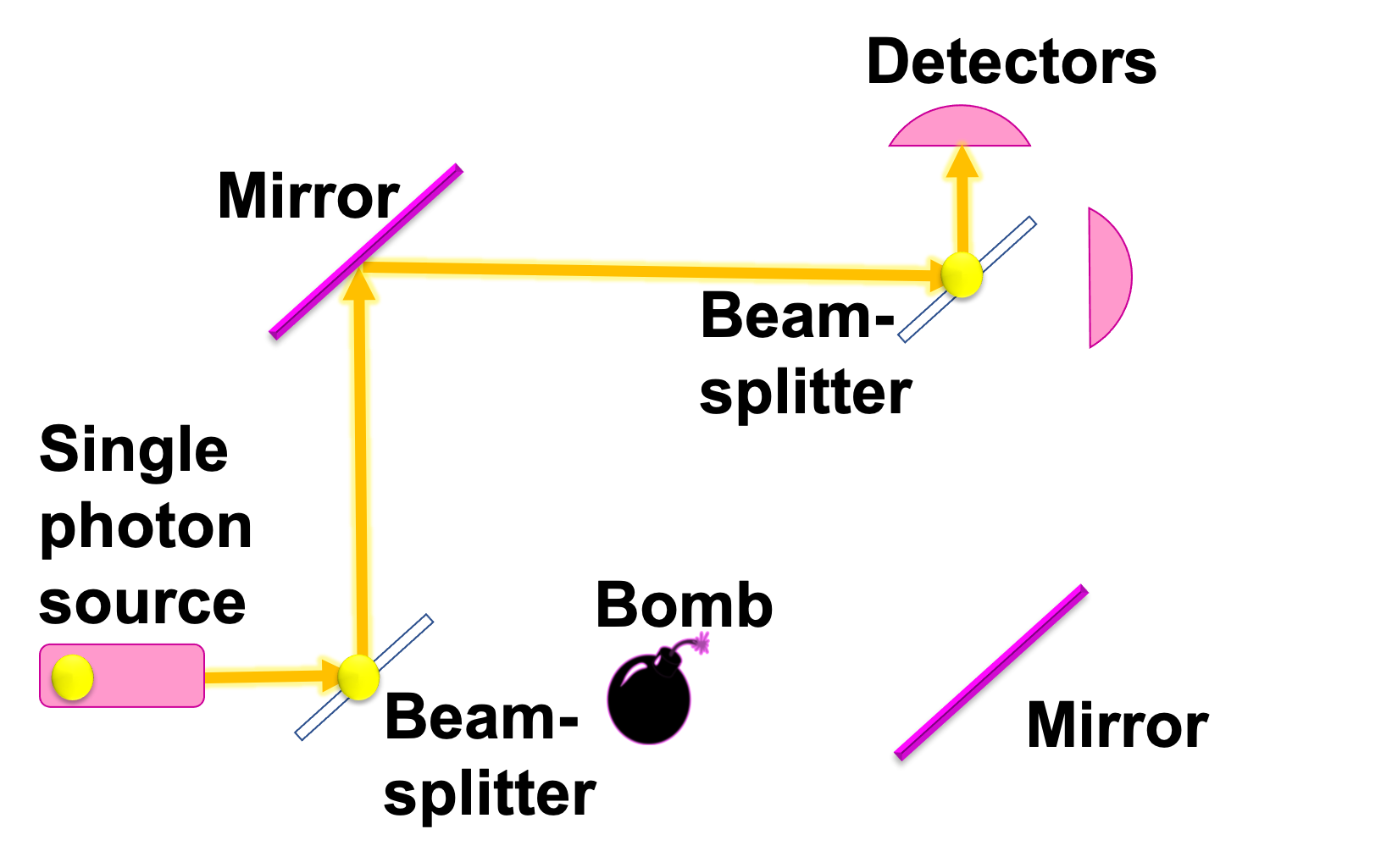}}
\caption{The quantum bomb tester thought experiment is depicted here for the case where the photon is projected into the path without the bomb, and is projected into the top detector, indicating that the bomb was present, without detonating it.}
\label{fig:6_bomb}
\end{figure}

\begin{figure}[htbp!]
\centerline{\includegraphics[width=0.5\textwidth]{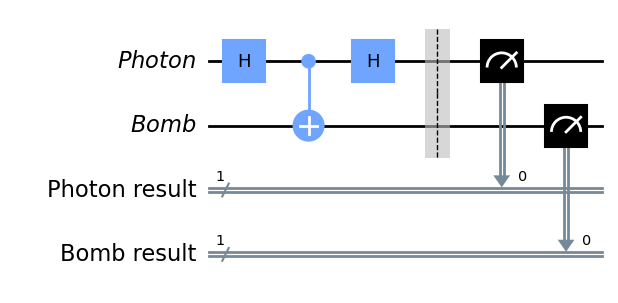}}
\caption{Bomb tester with bomb experiment}
\label{fig:bomb_tester_with_bomb}
\end{figure}

Imagine you are given a box and you cannot see what is inside. All you know is that it is either empty -- or it contains a highly sensitive bomb. If this bomb is hit by even a single photon, it will explode. Your challenge is to work out whether or not there is a bomb in the box, without exploding it.

Classically, there is no way to check if there is a bomb in the box without exploding it. As soon as we open the box to take a look, if the bomb is there it will be hit by a particle and explode.

Avshalom Elitzur and Lev Vaidman came up with a ``quantum bomb tester", which uses the quantum properties of photons to work out if a bomb is there without exploding it \cite{Elitzur1993}. They suggested putting the box into one path of the Mach-Zehnder interferometer, let's say it is the bottom path. If there is no bomb, then after passing through the empty box, the photon will then interfere with the top path and end up going straight through the second beam-splitter, as with the standard Mach-Zehnder interferometer setup. But if the bomb is there, the photon will be projected from the superposition to either the top or bottom path. If it is projected into the path with the bomb, then the bomb will explode. But if it is projected into the top path, it will then reach the second beam-splitter, and split into a superposition of being reflected and going straight through. So if we place detectors around the second beam-splitter, and detect the photon was reflected, we know for sure that the bomb is there, without setting it off! Now in this setup, there is only a 25$\%$ chance that we detect the bomb is there without setting it off. But other bomb testers were later proposed that allow us to detect the bomb with 100$\%$ probability \cite{Kwiat_bomb}.

To translate this setup to a quantum circuit, I use a qubit to represent the photon and a qubit to represent the bomb. As before, each beam-splitter is a Hadamard gate. 

After the first beam-splitter we may or many not have a bomb present, which effectively measures which path the photon has taken, causing the projection of the photon into one of the two paths. I will represent this bomb by a CNOT gate between the photon qubit and bomb qubit. If the CNOT gate is present (i.e. the bomb is in the path of the photon), then:
If the photon qubit is in the 0 state, nothing happens to the bomb qubit (it stays "unexploded" in the $\ket{0}$ state).
If the photon qubit is in the 1 state, then the bomb qubit is flipped to a $\ket{1}$, the exploded state.

Finally we pass the photon through another beam-splitter, which is another Hadamard gate on the photon qubit. Then we measure the photon's state: if it is a $\ket{0}$, then the CNOT may or may not have been there, so we do not know whether there was a bomb. If it is a $\ket{1}$, then we know for sure that the CNOT was there, so there was a bomb. We also need to measure our bomb qubit to check if it has exploded or not exploded (given by a $\ket{1}$ or a $\ket{0}$ respectively). If we manage to get the photon being a $\ket{1}$ and the bomb a $\ket{0}$ then we have success -- we have measured the bomb was there without exploding it. The full quantum circuit is shown in Figure $\ref{fig:bomb_tester_with_bomb}$.

\section{A Recipe for Resolutions}

There are some general principles which come up multiple times when analyzing paradoxical-seeming thought experiments and translating them to quantum circuits. These enable the apparent paradoxes to be resolved without contradictions: \\

\begin{enumerate}
    \item Treating detectors as quantum systems.
    \item Treating measurements as entangling CNOT gates.
    \item Treating the environment as a quantum system that induces decoherence.\\
\end{enumerate}

This approach of translating a complex setup, which includes detectors, into a quantum circuit, has helped clarify widely discussed thought experiments such as the quantum pigeonhole paradox and the Frauchiger-Renner paradox (\cite{aharonov2016quantum, frauchiger2018quantum, nurgalieva2022thought}). Additionally, it has been shown that a medium (such as a detector) able to couple with a quantum system must also itself have non-classical features \cite{marletto2022quantum}, reinforcing the need to treat detectors as non-classical. Similar principles are important in the recently proposed thought experiments for witnessing non-classicality in gravity and other systems \cite{marletto2017gravitationally, marletto2020witnessing}. 

Hence, the approach of translating quantum thought experiments to quantum circuits is not only a useful tool for understanding thought experiments and exploring the possibilities of quantum circuits on near-term devices, but is founded on principles that touch on deep ideas at the core of quantum science and technology. 

\section{Quantum bomb tester workshop example}

Here I will outline how the quantum bomb tester thought experiment can be used to give an introductory workshop to quantum science and technology, including how to code a quantum computer. 

\textbf{Motivation.} The motivation for basing an introductory quantum computing workshop on the quantum bomb tester thought experiment is that it has a compelling story, simple setup, foundational and technological implications, and a simple-to-code implementation using the Hadamard and CNOT gates. The implementation of a quantum circuit for the arbitrarily good quantum bomb tester uses more advanced quantum gates, and can be used as an extension or to provide a conceptual and coding challenge for more advanced participants.  

\textbf{Timing and format.} The workshop can fit into a 1-hour session, accessible to students as young as 11. The coding part of the workshop can be based on a Jupyter Notebook with sections to complete, with more or less code blocks filled in to adjust the difficulty level as appropriate for the age and experience of the participants. The Notebook can be run on a cloud service to remove the need for installing Python and Qiskit on personal computers, e.g. using the IBM Quantum Lab on the cloud or Google Colab notebooks on the cloud. The coding element could be done on a single demonstrator's computer if presenting to a large audience, or it could be done in pairs or small groups if there are laptops available. 

\textbf{Really? As young as 11??} In an ``Introduction to Quantum Computing" workshop for 11-to-14 year old girls, at the Oxford Physics Marie Curious event 2020, we gave participants a high-level overview of qubits and basic quantum gates before they successfully engaged with and completed Qiskit coding tasks. With sufficient pre-filled code in the workshop Notebook, even young students can engage with coding and understanding quantum circuits. 

\textbf{Resources and code.}
The Notebook can be based on my Qiskit code tutorial on the quantum bomb tester, which also has an explanatory video and blog post \cite{violaris2023quantum, violaris2023quantumcode}. Note that this content assumes basic knowledge of quantum circuits, which needs to be introduced separately in a workshop for new quantum learners. 

\textbf{Background.} I first turned this thought experiment into the quantum circuit presented in this paper when creating Qiskit workshops for the Oxford Quantum Information Society in 2019 -- 2020, in which I used the bomb tester as an extension activity. 

\subsection{Workshop Agenda}

\begin{enumerate}
\item \textbf{Setting the Stage:} Begin by explaining the quantum bomb tester problem. Emphasize the impossibility of detecting the bomb classically without it exploding, whereas quantum mechanics provides a resolution.

\item \textbf{Introduction to Mach-Zehnder Interferometer:} Explain the operation of the Mach-Zehnder interferometer without the presence of a bomb. Discuss how a beam-splitter induces a superposition of two paths for a photon and how the photon paths interfere constructively at the second beam-splitter to produce a single photon.

\item \textbf{The Effect of a Bomb:} Discuss the alteration in the interferometer's operation when a bomb is placed in one of its paths. Highlight the projection of the photon into either the path containing the bomb (resulting in detonation) or the path without the bomb. Explain how the different outcomes at the second beam-splitter, due to the absence of interference, can be used to successfully detect the bomb without exploding it!

\item \textbf{Mapping to Quantum Circuits:} Translate the interferometer setup to the setting of quantum bits (qubits) and quantum gates. Introduce the notion of qubits, the Hadamard gate, and the CNOT gate. Explain how the photon and bomb can be modeled as qubits and the interaction between them can be represented with quantum gates. 

\item \textbf{Running the Quantum Circuit:} Guide the participants to code the quantum circuits themselves and run them on a quantum simulator. Discuss the different outcomes for scenarios with and without the bomb.

\item \textbf{Interactive Game - Quantum Minesweeper:} Engage participants in a game of "Quantum Minesweeper" based on the more effective bomb tester. This could involve participants competing to get the highest score, by having the best success rate at determining the presence of a bomb without detonating it. 

\item \textbf{The Philosophy and the Technology:} Depending on the interests of the participants, you can discuss the philosophical implications of the quantum bomb tester (e.g. how Vaidman proposed it as evidence for the many-worlds theory of quantum mechanics) and the technological applications (e.g. interaction-free measurement being applied to imaging delicate objects, and counterfactual quantum communication and quantum computing protocols).  

\item \textbf{Extension - 100\% Effective Bomb Tester:} For participants with more experience, extend the workshop to explain the 100\% effective bomb tester. This forms the mechanism behind the Quantum Minesweeper game. The concept can first be explained using interferometers, and then participants can either be shown how to code it or be challenged to find a way to code it themselves.

\end{enumerate}

\section{Summary and conclusion}

I have explained how quantum thought experiments can be translated into quantum circuits, providing interesting activities for learners to code as quantum circuits for quantum computers, and to understand deep aspects of superposition, entanglement, measurement and decoherence. I have considered Schrödinger's cat, Wigner's friend, Deutsch's test of many-worlds, the double-slit, the delayed-choice quantum eraser, and the quantum bomb tester. I discussed some connecting factors that help transform quantum thought experiments to quantum circuits, namely by treating detectors as quantum systems and measurements as CNOT gates. This method elucidates subtle aspects which can make the thought experiments appear contradictory, resolving apparent paradoxes and showing that universal quantum theory is self-consistent. While the validity of universal quantum theory remains a topic of debate \cite{schlosshauer2013snapshot}, I have shown how quantum circuits are useful tools for concretely demonstrating and understanding the theory's implications. 

Next I explained how an introductory quantum workshop can be structured around the quantum bomb tester thought experiment, and made accessible to participants as young as 11. The workshop involves setting the scene of an engaging story with the context of the bomb-tester problem, and explaining how the bomb can be detected using an interferometer. Then qubits and the basic quantum gates, Hadamard and CNOT, are introduced by comparison with components of the interferometer. The bomb tester can then be coded as a quantum circuit e.g. with Qiskit, for participants to run and experiment with themselves. The ``Quantum Minesweeper" game that I created in the code tutorial \cite{violaris2023quantumcode} provides a fun way to interact with the bomb tester, and the improved bomb tester with up to 100$\%$ success provides an extension opportunity for advanced participants. The workshop, based on a blend of slide presentations and a Jupyter Notebook with an appropriate level of pre-filled code, facilitates the participants in coding their own quantum bomb tester and running quantum circuits on simulators. These could also be run on hardware, if access is available during the workshop. 

For a more comprehensive exploration of a broad range of quantum thought experiments translated into quantum circuits, I invite the reader to my ongoing Qiskit content series, comprising of videos, blog posts, and code tutorials \cite{violaris2023quantum}. This already includes the quantum bomb tester, and there is content in preparation on all of the thought experiments explained in this paper, which will include Jupyter Notebooks that learners can use to run the code themselves.

Another useful tool for playing with quantum thought experiments, especially those using the Mach-Zehnder interferometer and behaviour of photons, is the drag-and-drop Virtual Lab developed by Quantum Flytrap \cite{migdal2022visualizing}. There are many ready-made demonstrations in the Experimental Setups section, including a bomb tester and quantum eraser \cite{QuantumFlytrap2023}.

Quantum thought experiments began in the physics lab inside our heads. Now, quantum computers are putting opportunities to implement, simulate and get creative with these powerful research tools directly into the hands of quantum learners --- and the thought experiments in turn provide the perfect opportunity to get creative with quantum computing.

\bibliographystyle{IEEEtran}
\bibliography{references}  

\begin{thebibliography}{10}
\providecommand{\url}[1]{#1}
\csname url@samestyle\endcsname
\providecommand{\newblock}{\relax}
\providecommand{\bibinfo}[2]{#2}
\providecommand{\BIBentrySTDinterwordspacing}{\spaceskip=0pt\relax}
\providecommand{\BIBentryALTinterwordstretchfactor}{4}
\providecommand{\BIBentryALTinterwordspacing}{\spaceskip=\fontdimen2\font plus
\BIBentryALTinterwordstretchfactor\fontdimen3\font minus
  \fontdimen4\font\relax}
\providecommand{\BIBforeignlanguage}[2]{{%
\expandafter\ifx\csname l@#1\endcsname\relax
\typeout{** WARNING: IEEEtran.bst: No hyphenation pattern has been}%
\typeout{** loaded for the language `#1'. Using the pattern for}%
\typeout{** the default language instead.}%
\else
\language=\csname l@#1\endcsname
\fi
#2}}
\providecommand{\BIBdecl}{\relax}
\BIBdecl

\bibitem{Schrodinger1935}
E.~Schrödinger, ``Die gegenwärtige situation in der quantenmechanik,''
  \emph{Die Naturwissenschaften}, vol.~23, no.~48, pp. 807--812, 1935.

\bibitem{Wigner1961}
E.~P. Wigner, ``Remarks on the mind-body question,'' \emph{In Symmetries and
  Reflections: Scientific Essays}, 1961.

\bibitem{deutsch1985quantum}
D.~Deutsch, ``Quantum theory as a universal physical theory,''
  \emph{International Journal of Theoretical Physics}, vol.~24, pp. 1--41,
  1985.

\bibitem{Feynman1965}
R.~P. Feynman, R.~B. Leighton, and M.~Sands, \emph{The Feynman Lectures on
  Physics}.\hskip 1em plus 0.5em minus 0.4em\relax Addison-Wesley, 1965,
  vol.~3.

\bibitem{scully1982quantum}
M.~O. Scully and K.~Dr{\"u}hl, ``Quantum eraser: A proposed photon correlation
  experiment concerning observation and" delayed choice" in quantum
  mechanics,'' \emph{Physical Review A}, vol.~25, no.~4, p. 2208, 1982.

\bibitem{Kim2000}
Y.-H. Kim, R.~Yu, S.~P. Kulik, Y.~Shih, and M.~O. Scully, ``A delayed choice
  quantum eraser,'' \emph{Physical Review Letters}, vol.~84, no.~1, pp. 1--5,
  2000.

\bibitem{zurek1986frontiers}
W.~H. Zurek, ``Frontiers of nonequilibrium statistical physics,'' \emph{GT
  Moore, and MO Scully Plenum, New York}, 1986.

\bibitem{Leff2003}
H.~S. Leff and A.~F. Rex, \emph{Maxwell's Demon 2: Entropy, Classical and
  Quantum Information, Computing}.\hskip 1em plus 0.5em minus 0.4em\relax
  Institute of Physics Publishing, 2003.

\bibitem{deutsch1991quantum}
D.~Deutsch, ``Quantum mechanics near closed timelike lines,'' \emph{Physical
  Review D}, vol.~44, no.~10, p. 3197, 1991.

\bibitem{Nielsen2010}
M.~A. Nielsen and I.~L. Chuang, \emph{Quantum Computation and Quantum
  Information}.\hskip 1em plus 0.5em minus 0.4em\relax Cambridge University
  Press, 2010.

\bibitem{Qiskit}
{Qiskit contributors}, ``Qiskit: An open-source framework for quantum
  computing,'' 2023.

\bibitem{Elitzur1993}
A.~C. Elitzur and L.~Vaidman, ``Quantum mechanical interaction-free
  measurements,'' \emph{Foundations of Physics}, vol.~23, no.~7, pp. 987--997,
  1993.

\bibitem{zurek2003decoherence}
W.~H. Zurek, ``Decoherence, einselection, and the quantum origins of the
  classical,'' \emph{Reviews of modern physics}, vol.~75, no.~3, p. 715, 2003.

\bibitem{bild2023cat}
\BIBentryALTinterwordspacing
M.~Bild, M.~Fadel, Y.~Yang, U.~von Lüpke, P.~Martin, A.~Bruno, and Y.~Chu,
  ``Schrödinger cat states of a 16-microgram mechanical oscillator,''
  \emph{Science}, vol. 380, no. 6642, pp. 274--278, 2023. [Online]. Available:
  \url{https://www.science.org/doi/abs/10.1126/science.adf7553}
\BIBentrySTDinterwordspacing

\bibitem{Kwiat_bomb}
\BIBentryALTinterwordspacing
P.~Kwiat, H.~Weinfurter, T.~Herzog, A.~Zeilinger, and M.~A. Kasevich,
  ``Interaction-free measurement,'' \emph{Phys. Rev. Lett.}, vol.~74, pp.
  4763--4766, Jun 1995. [Online]. Available:
  \url{https://link.aps.org/doi/10.1103/PhysRevLett.74.4763}
\BIBentrySTDinterwordspacing

\bibitem{aharonov2016quantum}
Y.~Aharonov, F.~Colombo, S.~Popescu, I.~Sabadini, D.~C. Struppa, and
  J.~Tollaksen, ``Quantum violation of the pigeonhole principle and the nature
  of quantum correlations,'' \emph{Proceedings of the National Academy of
  Sciences}, vol. 113, no.~3, pp. 532--535, 2016.

\bibitem{frauchiger2018quantum}
D.~Frauchiger and R.~Renner, ``Quantum theory cannot consistently describe the
  use of itself,'' \emph{Nature communications}, vol.~9, no.~1, p. 3711, 2018.

\bibitem{nurgalieva2022thought}
N.~Nurgalieva, S.~Mathis, L.~del Rio, and R.~Renner, ``Thought experiments in a
  quantum computer,'' \emph{arXiv preprint arXiv:2209.06236}, 2022.

\bibitem{marletto2022quantum}
C.~Marletto and V.~Vedral, ``The quantum totalitarian property and exact
  symmetries,'' \emph{AVS Quantum Science}, vol.~4, no.~1, p. 015603, 2022.

\bibitem{marletto2017gravitationally}
------, ``Gravitationally induced entanglement between two massive particles is
  sufficient evidence of quantum effects in gravity,'' \emph{Physical Review
  Letters}, vol. 119, no.~24, p. 240402, 2017.

\bibitem{marletto2020witnessing}
------, ``Witnessing nonclassicality beyond quantum theory,'' \emph{Physical
  Review D}, vol. 102, no.~8, p. 086012, 2020.

\bibitem{violaris2023quantum}
\BIBentryALTinterwordspacing
M.~Violaris, ``Quantum paradoxes,'' 2023, [Online; accessed 1st June 2023].
  [Online]. Available:
  \url{https://www.youtube.com/playlist?list=PLOFEBzvs-VvoQP-EVyd5Di3UrPPc2YKIc}
\BIBentrySTDinterwordspacing

\bibitem{violaris2023quantumcode}
\BIBentryALTinterwordspacing
------, ``Quantum minesweeper code,'' 2023, [Online; accessed 1st June 2023].
  [Online]. Available:
  \url{https://github.com/maria-violaris/quantum-paradoxes/blob/main/quantum-minesweeper-code.ipynb}
\BIBentrySTDinterwordspacing

\bibitem{schlosshauer2013snapshot}
M.~Schlosshauer, J.~Kofler, and A.~Zeilinger, ``A snapshot of foundational
  attitudes toward quantum mechanics,'' \emph{Studies in History and Philosophy
  of Science Part B: Studies in History and Philosophy of Modern Physics},
  vol.~44, no.~3, pp. 222--230, 2013.

\bibitem{migdal2022visualizing}
P.~Migda{\l}, K.~Jankiewicz, P.~Grabarz, C.~Decaroli, and P.~Cochin,
  ``Visualizing quantum mechanics in an interactive simulation--virtual lab by
  quantum flytrap,'' \emph{Optical Engineering}, vol.~61, no.~8, pp.
  081\,808--081\,808, 2022.

\bibitem{QuantumFlytrap2023}
\BIBentryALTinterwordspacing
P.~Migdał and K.~Jankiewicz, ``Quantum flytrap: A no-code online laboratory
  for quantum experiments,'' Quantum Flytrap, 2023, available at:
  \url{https://quantumflytrap.com/virtual-lab}. [Online]. Available:
  \url{https://quantumflytrap.com/virtual-lab}
\BIBentrySTDinterwordspacing

\end{thebibliography}

\end{document}